\documentclass[11pt,a4paper]{article}
\usepackage{times,latexsym}
\usepackage{url}
\usepackage[T1]{fontenc}
\usepackage{amsmath,amssymb,amsfonts}
\usepackage{graphicx}
\usepackage{textcomp}
\usepackage{xcolor}
\usepackage{multirow}
\usepackage{multicol}
\usepackage{enumitem}
\usepackage{balance}
\usepackage{parallel,enumitem}
\usepackage{algorithmic}
\usepackage[linesnumbered,ruled]{algorithm2e}
\usepackage{bm}
\usepackage{booktabs}
\usepackage{url}
\usepackage[acceptedWithA]{tacl2021v1}

\usepackage{xspace,mfirstuc,tabulary}

\newcommand{\change}[1]{\textcolor{black}{{#1}}}
\newif\iftaclinstructions
\taclinstructionsfalse 
\iftaclinstructions
\renewcommand{\confidential}{}
\renewcommand{\anonsubtext}{(No author info supplied here, for consistency with
TACL-submission anonymization requirements)}
\newcommand{\instr}
\fi

\iftaclpubformat 

\else

\fi

\title{Tracking Brand-Associated Polarity-Bearing Topics in User Reviews}

\author{
  Runcong Zhao$^{1,2}$, Lin Gui$^1$, Hanqi Yan$^2$, Yulan He$^{1,2,3}$ 
  \\
  $^1$King's College London,  $^2$University of Warwick,  $^3$The Alan Turing Institute
  \\
  \texttt{runcong.zhao@warwick.ac.uk, yulan.he@kcl.ac.uk}
}

\date{}

\begin{document}
\maketitle
\begin{abstract}
Monitoring online customer reviews is important for business organisations to measure customer satisfaction and better manage their reputations. In this paper, we propose a novel dynamic Brand-Topic Model (dBTM) which is able to automatically detect and track brand-associated sentiment scores and polarity-bearing topics from product reviews organised in temporally-ordered time intervals. dBTM models the evolution of the latent brand polarity scores and the topic-word distributions over time by Gaussian state space models. It also incorporates a meta learning strategy to control the update of the topic-word distribution in each time interval in order to ensure smooth topic transitions and better brand score predictions. It has been evaluated on a dataset constructed from MakeupAlley reviews \change{and a hotel review dataset}. Experimental results show that dBTM outperforms a number of competitive baselines in brand ranking, achieving a good balance of topic coherence and uniqueness, and extracting well-separated polarity-bearing topics across time intervals\footnote{Data and code are available at \url{https://github.com/BLPXSPG/dBTM}.}. 
\end{abstract}

\section{Introduction}
With the increasing popularity of social media platforms, customers tend to share their personal experience towards products online. Tracking customer reviews online could help business organisations to measure customer satisfaction and better manage their reputations. Monitoring brand-associated topic changes in reviews can be done through the use of dynamic topic models~\cite{blei2006dynamic,wang2008continuous,dieng2019dynamic}. Approaches such as the dynamic Joint Sentiment-Topic (dJST) model~\cite{djst2014} are able to extract polarity-bearing topics evolved over time by assuming the dependency of the sentiment-topic-word distributions across time slices. They however require the incorporation of word prior polarity information and assume topics are associated with discrete polarity categories. Furthermore, they are not able to infer brand polarity scores directly. 

A recently proposed Brand-Topic Model (BTM) \cite{ZhaoGPH21} is able to automatically infer real-valued brand-associated sentiment scores from reviews and generate a set of sentiment-topics by gradually varying its associated sentiment scores from negative to positive. This allows users to detect, for example, strongly positive topics or slightly negative topics. BTM however assumes all documents are available prior to model learning and cannot track topic evolution and brand polarity changes over time.

\begin{figure*}
    \centering
    \includegraphics[width=1\linewidth]{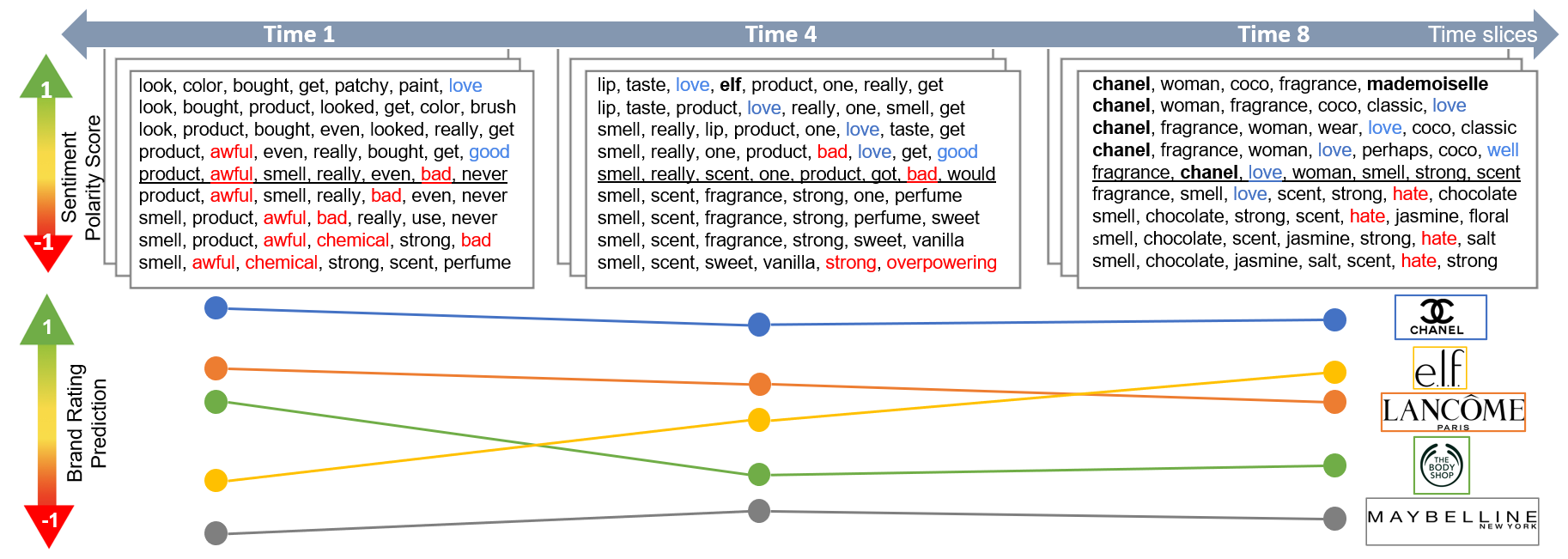}
    \caption{Brand-associated polarity-bearing topics tracking by our proposed model.  We show top words from an example topic extracted in time slice 1, 4 and 8 along the horizontal axis. In each time slice, we can see 
    a set of topics 
    generated by gradually varying their associated sentiment scores from -1 (\emph{negative}) to 1 (\emph{positive} along the vertical axis.  For easy inspection, positive words are highlighted in blue while negative ones in red. We can observe in Time 1, negative topics are mainly centred on the complaint of the chemical smell of a perfume, while positive topics are about the praise of the look of a product. From Time 1 to Time 8, we can also see the evolving aspects in negative topics moving from complaining about the strong chemical of perfume to overpowering sweet scent. In the lower part of the figure, we show the inferred polarity scores of three brands. For example, Channel is generally ranked higher than Lanc\^{o}me, which in turn scores higher than The Body Shop.}
    \label{fig:overview}
\end{figure*}

In this paper, we propose a novel framework inspired by Meta-Learning, which is widely used for distribution adaptation tasks \cite{SuoCZZ20}. When training the model on temporally-ordered documents divided into time slice, we assume that extracting polarity-bearing topics and inferring brand polarity scores in each time slice can be treated as a new sub-task and the goal of model learning is to learn to adapt the topic-word distributions associated with different brand polarity scores in a new time slice. We use BTM as the base model and store the parameters learned in a memory. At each time slice, we gauge model performance on a validation set based on the model-generated brand ranking results. The evaluation results are used for early stopping and dynamically initialising model parameters in the next time slice with meta learning. 
The resulting model is called dynamic Brand Topic Modelling (dBTM).

The final outcome from dBTM is illustrated in Figure \ref{fig:overview}, in which it can simultaneously track topic evolution and infer latent brand polarity score changes over time. Moreover, it also enables the generation of fine-grained polarity-bearing topics in each time slice by gradually varying brand polarity scores. In essence, we can observe topic transitions in two dimensions, either along a discrete time dimension, or along a continuous brand polarity score dimension.

We have evaluated dBTM on a review dataset constructed from MakeupAlley\footnote{ \url{https://www.makeupalley.com/}}, consisting of over 611K reviews spanning over 9 years, \change{and a hotel review dataset sampled from HotelRec \cite{hotelrec}, containing reviews of the most popular 25 hotels over 7 years}. We compare its performance with a number of competitive baselines and observe that it generates better brand ranking results, predicts more accurate brand score time series, 
and produces well-separated polarity-bearing topics with more balanced topic coherence and diversity. More interestingly, we have evaluated dBTM in a more difficult setup, where the supervised label information, i.e., review ratings, is only supplied in the first time slice and afterwards, dBTM is trained in an unsupervised way without the use of review ratings. dBTM under such a setting can still produce brand ranking results across time slices more accurately compared to baselines trained under the supervised setting. 
This is a desirable property as dBTM, initially trained on a small set of labelled data, can self-adapt its parameters with streaming data in an unsupervised way.

Our contributions are three-fold:
\begin{itemize}
\item  We propose a new model, called dBTM, built on the Gaussian state space model with meta learning for dynamic brand topic and polarity score tracking;
\item We develop a novel meta learning strategy to dynamically initialise the model parameters at each time slice in order to better capture rating score changes, which in turn generates topics with a better overall quality;

\item  Our experimental results show that dBTM trained with the supervision of review ratings at the initial time slice, can self-adapt its parameters with streaming data in an unsupervised way and yet still achieve better brand ranking results compared to supervised baselines.

\end{itemize}

\section{Related Work}

Our work is related to the following research:
\subsection{Dynamic Topic models}
Topic models such as the Latent Dirichlet Allocation (LDA) model \cite{BleiNJ03} is one of the most successful approaches for the statistical analysis of document collections. 
Dynamic topic models aim to analyse the temporal evolution of topics in large document collections over time. 
Early approaches built on LDA 
include the dynamic topic model (DTM)~\cite{blei2006dynamic}, 
which use the Kalman filter to model the transition of topics across time, and the continuous time dynamic topic model~\cite{wang2008continuous} which replaced the discrete state space model of the DTM with its continuous generalisation. 
More recently, 
DTM is combined with word embeddings in order to generate more diverse and coherent topics in document streams \cite{dieng2019dynamic}. 

Apart from the commonly used LDA, Poisson factorisation can also be used for topic modelling, in which it factorises a document-word count matrix into a product of 
a document-topic matrix and a topic-word matrix. 
It can be extended to analyse 
sequential count vectors such as a document corpus which contains a single word count matrix with one column per time interval, 
by capturing dependence among time steps by a Kalman filter \cite{charlin2015dynamic}, 
neural networks \cite{gong2017deep}, or by extending a Poisson distribution on the document-word counts as a non-homogeneous Poisson process over time \cite{hosseini2018recurrent}. 

While the aforementioned models are typically used in the unsupervised setting, the Joint Sentiment-Topic (JST) model~\cite{lin2009joint,lin2012weakly} incorporated the polarity word prior into model learning, which enables the extraction of topics grouped under different sentiment categories. JST is later extended into a dynamic counterpart, called dJST, which tracks both topic and sentiment shifts over time \cite{djst2014} by assuming 
that the sentiment-topic word distribution at the current time is generated from the Dirichlet distribution parameterised by the sentiment-topic word distributions at previous time intervals.

\subsection{Market/Brand Topic Analysis}
LDA and its variants have been explored for marketing research. Examples include user interests detection by analysing consumer purchase behaviour \cite{dstr2017,sun2021}, the tracking of the competitors in the luxury market among given brands by mining the Twitter data \cite{dtm2015}, and identify emerging app issues from user reviews \cite{tour2021}. Matrix factorization which is able to extract the global information is also used to be applied in product recommendation \cite{zhourecom2020}, review summarization \cite{cuisum2021}.
The interaction between topics and polarities can be modelled by the incorporation of approximations by sampling based methods \cite{lin2009joint} with sentiment prior knowledge such as sentiment lexicon~\cite{lin2012weakly}. But such prior knowledge would be highly domain specific. Seed words with known polarities or seed words generated by morphological information \cite{unsupas2010} is another common method to get topic polarity. But those methods are focused on analysing the polarity of existing topics. 
More recently, the Brand-Topic Model built on Poisson factorisation was proposed \cite{ZhaoGPH21}, which can infer brand polarity scores and generate fine-grained polarity-bearing topics. \change{The detailed description of BTM can be found at Section 3.}

\subsection{Meta Learning}
Meta learning, or learning to learn, 
can be broadly categorised into metric-based learning and optimisation-based learning. 
Metric-based learning aims to learn a distance function between training instances so that it can classify a test instance by comparing it with the training instances in the learned embedding space \cite{SungYZXTH18}. 
Optimisation-based learning 
usually splits the labelled samples into \emph{training} and \emph{validation} sets. The basic idea is to fine-tune the parameters on the \emph{training} set to obtain the updated parameters, which are then evaluated on the \emph{validation} set to get the error which is converted as a loss value for optimising the original parameters  \cite{FinnAL17,JamalQ19}. 
Meta learning has been explored in many tasks, including text classification \cite{GengLLSZ20},  
topic modelling \cite{TopicOcean}, 
knowledge representation \cite{ZhengYG021}, recommender systems \cite{Neupane21,DongYYXZ20,Lu0S20} and event detection \cite{DengZKZZC20}. Especially, the meta learning based methods have achieved significant successes in distribution adaptation \cite{SuoCZZ20,YuGHZLO021}. 
 We propose a meta learning strategy here to learn how to automatically initialise model parameters in each time slice. 

\section{Preliminary: Brand Topic Model}

The Brand-Topic Model (BTM) \cite{ZhaoGPH21}, as shown in the middle part of Figure~\ref{fig:architecture}, is trained on review documents paired with their document-level sentiment class labels (e.g., `\emph{Positive}', `\emph{Negative}' and `\emph{Neutral}'). It can automatically infer real-valued brand-associated polarity scores and generate fine-grained sentiment-topics in which a continuous change of words under a certain topic can be observed with a gradual change of its associated sentiment. It was partly inspired by the Text-Based Ideal Point (TBIP) model \cite{tbip}, which aims to model the generation of text via Poisson factorisation. 
In particular, for the input bag of words data, the count for term $v$ in document $d$ is formulated as term count $c_{dv}$, which is assumed to be sampled from a Poisson distribution $c_{dv} \sim \mbox{Poisson}(\lambda_{dv})$ where the rate parameter $\lambda_{dv}$ can be factorised as:

\begin{equation}
    { \lambda}_{dv} = \sum_k{\theta_{dk}\beta_{kv}}\label{eq:poisson-factorisation}
\end{equation}

Here, $\theta_{dk}$ denotes the per-document topic intensity, $\beta_{kv}$ represents the topic-word distribution. We have $\bm{\theta}\in\mathbb{R}_+^{D\times K}$, $\bm{\beta}\in\mathbb{R}_+^{K\times V}$, where $D$ is the total number of documents in a corpus, $K$ is the topic number, and $V$ is the vocabulary size.
Then, brand-polarity score $x_{b_{d}}$ and topic-word offset $\eta_{kv}$ are added to the model:

\begin{equation}
    {\lambda}_{dv} = \sum_k{\theta_{dk}\exp(\log\beta_{kv} + x_{b_{d}}\eta_{kv})}\label{eq:BTM-basic}
\end{equation}

$x_{b_{d}}$ is the brand polarity score for document $d$ of brand $b$ and 
we have $\bm{\eta}\in\mathbb{R}_+^{K\times V}$,  $\bm{x}\in\mathbb{R}$. \change{The model normalised brand polarity assignment to [-1,1] in its output for demonstration purposes}.

The intuition behind the above formulation is that the latent variable $x_{b_{d}}$ which captures the brand polarity score can be either positive or negative. If a word tends to frequently occur in reviews with positive polarities, but the polarity score of the current brand is negative, then the occurrence count of such a word would be reduced by making $x_{b_{d}}$ and $\eta_{kv}$ to have opposite signs. 

A Gamma prior is placed on $\bm{\theta}$ and $\bm{\beta}$, with $a,b,c,d$ being hyper-parameters, while a normal prior is placed over the brand polarity score $\bm{x}$ and the topic-word count offset $\bm{\eta}$. 
\begin{equation}
\begin{gathered}
    \theta_{dk} \sim \mbox{Gamma}(a,b), \quad
    \beta_{kv} \sim \mbox{Gamma}(c,d), \nonumber \\
     x_{b_d} \sim \mathcal{N}(0,1), \quad
    \eta_{kv} \sim \mathcal{N}(0,\bm{I}). \nonumber
\end{gathered}
\end{equation}

BTM makes use of Gumbel-Softmax \citep{gs2017} to construct document features for sentiment classification. \change{This is because directly sampling word counts from the Poisson distribution is not differentiable. Gumbel-Softmax, which is a gradient estimator with the reparameterization trick, is used to enable back-propagation of gradients.} More details can be found in \cite{ZhaoGPH21}.

\section{Dynamic Brand Topic Model (dBTM)}

\begin{figure}[htb]
\centering
\includegraphics[width=0.95\linewidth]{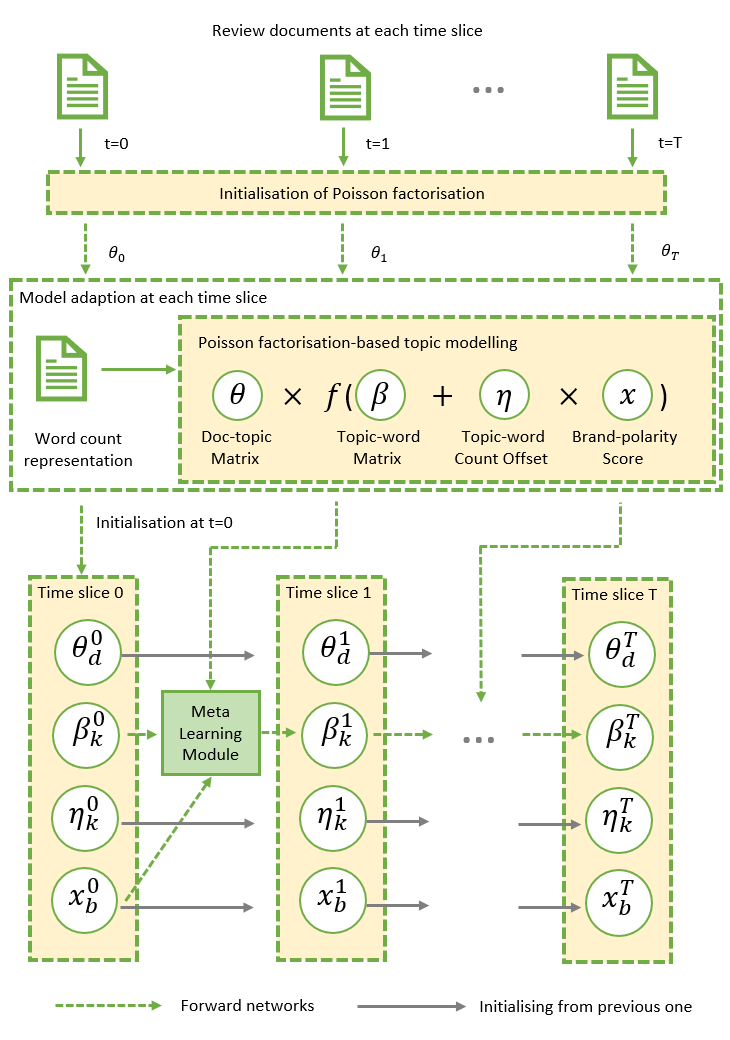}
\caption{The overall architecture of the dynamic Brand-Topic Model (dBTM), which extends the Brand-Topic Model (BTM) shown in the upper box to deal with streaming documents. In particular, at time slice $t$, the document-topic distribution $\bm{\theta}^t$ is initialised by a vanilla Poisson factorisation model, the evolution of the latent brand-associated polarity scores $\bm{x}^t$ and the polarity-associated topic-word offset $\bm{\eta}^t$ is modelled by two separate Gaussian state space models. The topic-word distribution $\bm{\beta}^t$ has its prior set based on the trend of the model performance on brand ranking results in the previous two time slices. Lines coloured in grey indicate parameters are linked by Gaussian state space models, while those coloured in green indicate forward calculations.}
\label{fig:architecture}
\end{figure}

To track brand-associated topic dynamics in customer reviews, we split the documents into time slices where the time period of each slice can be set arbitrarily at, e.g. a week, a month, or a year. In each time slice, we have a stream of $M$ documents $\{d_1,\cdots,d_M\}$ ordered by their publication timestamps. A document $d$ at time slice $t$ is input as a Bag-of-Words (BoW) representation. 
We extend BTM to deal with streaming documents by assuming that documents at the current time slice are influenced by documents at past. The resulting model is called dynamic Brand-Topic Model (dBTM) with its architecture illustrated in Figure~\ref{fig:architecture}. 

\subsection{Initialisation}

In the original BTM model, the latent variables to be inferred include the document-topic distribution $\bm{\theta}$, topic-word distribution $\bm{\beta}$, the brand-associated polarity score $\bm{x}$, and the polarity-associated topic-word offset $\bm{\eta}$. At time slice 0, we represent all documents in this slice as a document-word count matrix. We then perform Poisson factorisation \change{with coordinate-ascent variational inference \cite{hpf2015}} to derive $\bm{\theta}$ and $\bm{\beta}$ (see Eq. (\ref{eq:poisson-factorisation})). The topic-word count offset $\bm{\eta}$ and the brand polarity score $\bm{x}$ are sampled from a standard normal distribution.

\subsection{State-Space Model}
At time slice $t$, we can model the evolution of the latent brand-associated polarity scores $\bm{x}^t$ and the polarity-associated topic-word offset $\bm{\eta}^t$ over time by a Gaussian state space model:

{
\begin{align}
    \bm{x}^t | \bm{x}^{t-1} \sim \mathcal{N}(\bm{x}^{t-1}, \sigma_x^2\bm{I}) \\
    \bm{\eta}^t | \bm{\eta}^{t-1} \sim \mathcal{N}(\bm{\eta}^{t-1}, \sigma_\eta^2\bm{I}) 
\end{align}}

For the topic-word distribution $\bm{\beta}$, a similar Gaussian state-space model is adopted except that log-normal distribution is used:
\begin{equation}
    \bm{\beta}^t | \bm{\beta}^{t-1} \sim \mathcal{LN}(\bm{\beta}^{t-1}, \sigma_{\beta}^2\bm{I}) \label{eq:beta-evolution}
\end{equation}

While topic-word distribution could be inherited from previous time slice, the document-topic distribution $\bm{\theta}^t$ needs to be re-initialised at the start of each time slice since there is a different set of documents at each time slice. We propose to run a simple Poisson factorisation to derive the initial values of $\bm{\theta}_{(p)}^t$ before we do the model adaption at each time slice:
Here, the topic-word distribution in the previous time slice $\bm{\beta}_{(p)}^{t-1}$ becomes the prior of the topic-word distribution in the current time slice $\bm{\beta}_{(p)}^{t}$ as defined in Eq.~(\ref{eq:beta-evolution}). We use the subscript $(p)$ to denote that the parameters are derived in the Poisson factorisation initialisation stage at the start of each time slice.

Essentially at each time slice $t$, we initialise the document-topic distribution $\bm{\theta}^t$ of the BTM model as $\bm{\theta}_{(p)}^t$ which is obtained by performing Poisson factorisation on the document-word count matrix in $t$. For the topic-word distribution, within BTM, we can set $\bm{\beta}^{t}$ to be inherited from $\bm{\beta}^{t-1}$ as defined in Eq. (\ref{eq:beta-evolution}), but additionally, we also have $\bm{\beta}_{(p)}^{t}$ which is obtained by directly performing Poisson factorisation of the document-word count matrix in the current time slice. In what follows, we will present how we initialise the value of $\bm{\beta}^{t}$ through meta learning.

\subsection{Meta Learning}
We notice that although parameters in each time interval are linked with parameters in the previous time interval by Gaussian state-space models, the results generated at each time interval are not stable. Inspired by meta learning, we consider latent brand score prediction and sentiment topic extraction at each time interval as a new sub-task, and propose a learning strategy to dynamically initialise model parameters in each interval based on the brand rating prediction performance on the validation set of the previous interval.
In particular, we set aside 10\% of the training data in each time interval as the validation set and compare the model-inferred brand ranking result with the gold standard one using the Spearman's rank correlation coefficient. By default, 
the topic-word distribution in the current interval, $\bm{\beta}^{t}$, would have its prior set to 
$\bm{\beta}^{t-1}$ learned in the previous time interval. 
However, if the brand ranking result in the previous interval is poor, then $\bm{\beta}^{t}$ would be initialised as a weighted interpolation of $\bm{\beta}^{t-1}$ and the topic-word distribution obtained from the Poisson factorisation initialisation stage in the current interval $\bm{\beta}_{(p)}^{t}$. 
The rationale is that if the model performs poorly in the previous interval, then its learned topic-word distribution should have less impact on the parameters in the current interval. 
More concretely, we first evaluate the brand ranking result returned by the model at time slice $t-1$ on the validation set at $t-1$:

\begin{algorithm*}
 \algsetup{linenosize=\tiny}
  \scriptsize
    \SetKwInOut{Input}{Input}
    \SetKwInOut{Output}{Output}
    \Input{Number of topics $K$, number of brands $B$, time slice $t\in\{0,1,2,\cdots,T\}$, a stream of document-word count matrix $\bm{C}=\{\bm{c}^0,\cdots,\bm{c}^T\}$}
    \Output{Document-topic intensity $\{\bm{\theta}^t\}_{t=0}^T$, Topic-word matrix $\{\bm{\beta}^t\}_{t=0}^T$, brand scores $\{x_1^t,\cdots,x_B^t\}_{t=0}^T$}
    
    \underline{Initialisation:}\\
    Initialise $\bm{\theta}^0, \bm{\beta}^0$ by Poisson factorisation $\bm{c}^0 \sim \mbox{Poisson}(\bm{\theta}^0\bm{\beta}^0), \quad\quad s
     \bm{x}^0 \sim \mathcal{N}(0,\bm{I}), \quad\quad
    \bm{\eta}^0 \sim \mathcal{N}(0,\Sigma_{\eta}^2\bm{I}), \quad\quad
    \gamma^0 = 0$\\
    Update parameters by minimising the loss defined in Eq. \ref{eq:totalLoss}\\
        Derive the brand ranking $\hat{\bm{r}}^{0}$ on the validation set based on the inferred brand polarity score $\{x_b^{0}\}_{b=1}^B$\\
        Calculate the Spearman's rank correlation coefficient $\rho^{0} = \mbox{SpearmanRank}(\hat{\bm{r}}^{0}, \bm{r}^{0})$ \\
        Set the weight $\gamma^{1} = \mbox{max}\big(0.05, \Phi_{z_{\rho^{0}}}(0)\big)$ \\
    \underline{Training:}\\
    \For{$t=1$ \KwTo $T$}{
        \underline{Pre-training:} 
        $\bm{c}^t \sim \mbox{Poisson}(\bm{\theta}_{(p)}^t\bm{\beta}_{(p)}^t)$\\
        \underline{Per epoch initialisation:}\\
        $\bm{x}^t \sim \mathcal{N}(\bm{x}^{t-1}, \sigma_x^2\bm{I}), \quad
        \bm{\eta}^t \sim \mathcal{N}(\bm{\eta}^{t-1}, \sigma_\eta^2\bm{I}), \quad
        \bm{\theta}^t = \bm{\theta}_{(p)}^t, \quad
         \bm{\beta}^{t} = (1 - \gamma^{t}) \bm{\beta}^{t-1}  +  \gamma^{t} \bm{\beta}_{(p)}^{t}$\\
      \For{$i = 0$ \KwTo maximum iterations}{
      Update parameters by minimising the loss defined in Eq. \ref{eq:totalLoss}\\
      \If{checkpoint}{
        Derive the brand ranking $\hat{\bm{r}}^{t}$ on the validation set based on the inferred brand polarity score $\{x_b^{t}\}_{b=1}^B$\\
        Calculate the Spearman's rank correlation coefficient $\rho^{t} = \mbox{SpearmanRank}(\hat{\bm{r}}^{t}, \bm{r}^{t})$ \\
        \If{($\Phi_{z_{\rho^{t}}}(\rho^{t-1}) > 0.95$)}{break \emph{\# Null hypothesis is rejected by the upper quartile according to Eq. \ref{eq:distribution}}} 
        Set the weight   $\gamma^{t+1} = \mbox{max}\big(0.05,\Phi_{z_{\rho^{t}}}(\rho^{t-1})\big)$
         \\}
      }
    }
    \caption{Training procedure of dBTM}\label{alg:dBTM}
\end{algorithm*}

\begin{equation}
    \rho^{t-1} = \mbox{SpearmanRank}(\hat{\bm{r}}^{t-1}, \bm{r}^{t-1})
\end{equation}

\noindent where $\hat{\bm{r}}^{t-1}$ denotes the derived brand ranking result based on the model predicted latent brand polarity scores, $\hat{\bm{x}}^{t-1}$, at time slice $t-1$, $\bm{r}^{t-1}$ is the gold-standard brand ranking, and $\rho^{t-1}$ is the Spearman's rank correlation coefficient. To check if the brand ranking result gets worse or not, we compare it with the brand ranking evaluation result, $\rho^{t-2}$, in the earlier interval. In particular, we first take Fisher's z-transformation $z_{\rho^{t-1}}$ of $\rho^{t-1}$, which is assumed following a Gaussian distribution:
\begin{equation}
z_{\rho^{t-1}} \sim \mathcal{N} \left({ \mbox{ln}(\frac{1+\rho^{t-1}}{1-\rho^{t-1}})^{0.5},\frac{1}{B-3}}\right)
\label{eq:distribution}
\end{equation}
\noindent where $B$ denotes the total number of brands. Then we compute the Cumulative Distribution Function (CDF) of the above normal distribution, denoted as $\Phi_{z_{\rho^{t-1}}}$, and calculate $\Phi_{z_{\rho^{t-1}}}(\rho^{t-2})$, which essentially returns $\mbox{Pr}(z_{\rho^{t-1}} \leq \rho^{t-2})$. Lower value of $\Phi_{z_{\rho^{t-1}}}(\rho^{t-2})$ indicates that the model at $t-1$ generates a better brand rank result than that in the previous time slice $t-2$. This is equivalent to performing a hypothesis test in which we compare the rank evaluation result $\rho^{t-1}$ with $\rho^{t-2}$ to test if the model at $t-1$ performs better than that at $t-2$. The hypothesis testing result can be used to set the weight $\gamma^{t}$ to determine how to initialise the topic-word distribution at $t$, $\bm{\beta}^{t}$:
\begin{eqnarray}
 \gamma^{t} &= \mbox{max}\big(0.05,\Phi_{z_{\rho^{t-1}}}(\rho^{t-2})\big)\label{eq:gamma}\\
  \bm{\beta}^{t} &= (1 - \gamma^{t}) \bm{\beta}^{t-1}  + \gamma^{t} \bm{\beta}_{(p)}^{t}\label{eq:betaUpdate}
\end{eqnarray}
The above equations state that if the model trained at $t-1$ generates a better brand ranking result than that in the previous time slice significantly ($p$-value $>0.05$), then we are more confident to initialise $\bm{\beta}^{t}$ largely based on $\bm{\beta}^{t-1}$ according to the estimated probability of $\mbox{Pr}(z_{\rho^{t-1}} > \rho^{t-2}) = 1 - \gamma^{t}$. 
we will have to re-initialise $\bm{\beta}^{t}$ mostly based on the topic-word distribution obtained from the Poisson factorisation initialisation stage in the current interval $\bm{\beta}_{(p)}^{t}$.

\subsection{Parameter Inference}

We use the mean-field variational distribution to approximate the posterior distribution of latent variables, $\bm{\theta},\bm{\beta},\bm{\eta},\bm{x}$, given the observed document-word count data $\bm{c}$ by maximising the Evidence Lower-Bound (ELBO):
\begin{equation}
\begin{split}
    \mathcal{L}_{\mbox{ELBO}} = \mathbb{E}_{q_{\phi}}[\log\ p(\bm{\theta},\bm{\beta},\bm{\eta},\bm{x})] +  \\
    \log\ p(\bm{c}|\bm{\theta},\bm{\beta},\bm{\eta},\bm{x}) -  \log\ q_{\phi}(\bm{\theta},\bm{\beta},\bm{\eta},\bm{x})]
\end{split}
\end{equation}
where 
\begin{equation}
    q_{\phi}(\bm{\theta},\bm{\beta},\bm{\eta},\bm{x}) = \prod_{d,k,b}q(\theta_d)q(\beta_k)q(\eta_k)q(x_b)
\end{equation}

In addition, for each document $d$, we construct its representation $z_d$ by sampling word counts using Gumbel softmax from the aforementioned learned parameters, which is fed to a sentiment classifier to predict a class distribution $\hat{y}_d$. 
We also perform adversarial learning by inverting the sign of the inferred polarity score of the brand associated with document $d$ and produce the adversarial representation $\tilde{z}_d$. This is also fed to the same sentiment classifier which generates another predicted class distribution $\tilde{y}_d$. We train the model by minimising the Wasserstein distance between the prediction and the actual class distributions. The final loss function is the combination of the ELBO and the Wasserstein distance losses:
\begin{equation}
\begin{split}
    \mathcal{L} = -\mathcal{L}_{\mbox{ELBO}} + \frac{1}{M}\sum_{d=1}^M\big(\mathcal{L}_{\mbox{WD}}(\hat{y}_d, y_d) \\
    + \mathcal{L}_{\mbox{WD}}(\tilde{y}_d, \bar{y}_d)\big)\label{eq:totalLoss}
\end{split}
\end{equation}
where $\mathcal{L}_{\mbox{WD}}(\cdot)$ denotes the Wasserstein distance, $y_d$ is the gold-standard class distribution and $\bar{y}_d$ is the class distribution derived from the inverted document rating. By inverting the document rating, we essentially balance the document rating distributions that for each positive document, we also create a synthetic negative document, and vice versa.

\section{Experimental Setup}

\paragraph{\textbf{Datasets}}
\change{Popular datasets such as Yelp and Amazon products \cite{amazondata}  
and Multi-Domain Sentiment dataset \cite{blitzer2007biographies} 
are constructed by randomly selecting reviews from Amazon or Yelp without considering their distributions over various brands and across different time periods. 
Therefore, }
we construct our own dataset by crawling reviews from top 25 brands from MakeupAlley, 
a review website on beauty products. Each review is accompanied with a rating score, product type, brand and post time. 
\change{We consider reviews with the ratings of 1 and 2 as the negative class, those with the rating of 3 as the neutral class, and the remaining with the ratings of 4 and 5 as the positive class, following the label setting in BTM.} The entire dataset contains 611,128 reviews spanning over 9 years (2005 to 2013). 
We treat each year as a time slice and split reviews into 9 time slices.  
The average review length is 123 words. 
\change{Besides the MakeupAlley-Beauty, we also run our experiments on HotelRec \cite{hotelrec}, 
by selecting reviews from the top 25 hotels over 7 years (2012 to 2018)}.
The statistics of our datasets are shown in Table~\ref{tab:dataset_statistics}. 
It can be observed that the dataset is imbalanced with positive reviews being over triple the size of negative ones \change{for MakeupAlley-Beauty and nearly 10 times for HotelRec}.

\begin{table}[h]
\centering
\resizebox{\columnwidth}{!}{
\begin{tabular}{@{}lr@{}}
    \toprule
    \textbf{Dataset}  & \textbf{MakeupAlley-Beauty Reviews} \\
    \midrule
    No. of documents per class   &   \\
    $\ \ \ $ Neg / Neu / Pos   & 114,837 / 88,710 / 407,581  \\
    No. of brands & 25 \\
    Total no. of documents & 611,128  \\
    No. of time Slices & 9\\
    Average review length (\#words) & 123 \\
    Average no. of documents per slice & $\sim 68k$ \\
    Vocabulary size   & $\sim 4500$ \\ \midrule
    \change{\textbf{Dataset}}  & \change{\textbf{HotelRec Reviews}} \\ \midrule
    No. of documents per class   &   \\
    $\ \ \ $ Neg / Neu / Pos   & 14,600 / 20,629 / 150,265  \\
    No. of hotels & 25 \\
    Total no. of documents & 185,496  \\
    No. of time Slices & 7\\
    Average review length (\#words) & 204 \\
    Average no. of documents per slice & $\sim 26k$ \\
    Vocabulary size   & $\sim 7000$ \\
    \bottomrule
\end{tabular}
}
\caption{Dataset statistics of the reviews.}
\label{tab:dataset_statistics}
\end{table}

\begin{table*}[!t]
\centering

\resizebox{1.6\columnwidth}{!}{
\begin{tabular}{ccccccccccccccc}
\toprule
\multirow{2}{*}{Time Slice} & \multicolumn{2}{c}{dJST} & \multicolumn{2}{c}{TBIP} & \multicolumn{2}{c}{BTM}  & \multicolumn{2}{c}{O-dBTM} & \multicolumn{2}{c}{dBTM} \\
\cmidrule(lr){2-3} \cmidrule(lr){4-5} \cmidrule(lr){6-7} \cmidrule(lr){8-9} \cmidrule(lr){10-11} \cmidrule(lr){12-13} 
                                 & Corr       & p-value     & Corr       & p-value     & Corr           & p-value & Corr           & p-value         & Corr            & p-value  &  \\ \midrule
\multicolumn{11}{c}{\textit{MakeupAlley-Beauty}} \\ \midrule
1                                  & -0.249     & 0.230       & -0.567     & 0.003       & \textbf{0.552} & 0.004            & 0.454           & 0.023    & 0.402          & 0.046   \\
2                                  & -0.437     & 0.029       & \textbf{0.527}      & 0.007       & 0.488 & 0.013              & 0.459           & 0.021    & 0.438          & 0.029   \\
3                                  & -0.327     & 0.111       & -0.543     & 0.005       & -0.384         & 0.058           & 0.504           & 0.010    & \textbf{0.523} & 0.007   \\
4                                  & -0.127     & 0.545       & -0.431     & 0.032       & -0.428         & 0.033            & 0.448           & 0.025    & \textbf{0.453} & 0.023   \\
5                                  & 0.112      & 0.596       & -0.347     & 0.089       & 0.402          & 0.047              & \textbf{0.438}  & 0.028    & 0.394          & 0.051   \\
6                                  & -0.118     & 0.573       & -0.392     & 0.053       & 0.432          & 0.031           & 0.402           & 0.047    & \textbf{0.433} & 0.031   \\
7                                  & -0.203     & 0.330       & 0.400      & 0.048       & \textbf{0.417} & 0.038            & 0.400           & 0.048    & 0.402          & 0.047   \\
8                                  & -0.552      & 0.004       & 0.348      & 0.089       & 0.363          & 0.074            & 0.359           & 0.078    & \textbf{0.364} & 0.074   \\ \midrule
\multicolumn{11}{c}{\change{\textit{HotelRec}}} \\ \midrule
1 & 0.097  & 0.645 & 0.121          & 0.565 & -0.508 & 0.009 & \textbf{0.356} & 0.081 & 0.285          & 0.168 \\
2 & -0.242 & 0.244 & \textbf{0.443} & 0.027 & -0.337 & 0.100 & 0.196          & 0.347 & 0.382          & 0.059 \\
3 & -0.112 & 0.596 & -0.392         & 0.053 & 0.318  & 0.121 & \textbf{0.419} & 0.037 & 0.355          & 0.082 \\
4 & -0.362 & 0.076 & 0.276          & 0.181 & 0.301  & 0.144 & \textbf{0.349} & 0.087 & 0.315          & 0.126 \\
5 & -0.045 & 0.829 & 0.292          & 0.156 & 0.225  & 0.279 & 0.323          & 0.115 & \textbf{0.364} & 0.074 \\
6 & 0.222  & 0.285 & 0.298          & 0.148 & 0.306  & 0.137 & 0.294          & 0.154 & \textbf{0.312} & 0.130 \\

\bottomrule
\end{tabular}
}
\caption{
Brand ranking results generated by various models trained on time slice $t$ and tested on time slice $t+1$. We report the correlation coefficients \textit{corr} and its associated two-sided $p$-values.}
\label{tab:brand-ranking}
\end{table*}

\paragraph{\textbf{Models for Comparison}}
We conduct experiments using the following models:\\
\begin{itemize}
    \item \underline{Dynamic Joint Sentiment-Topic (dJST) model} \cite{djst2014}, built on LDA, can detect and track polarity-bearing topics from text with the word prior sentiment knowledge incorporated. In our experiments, the MPQA subjectivity lexicon\footnote{\url{https://mpqa.cs.pitt.edu/lexicons/}} is used to derive the word prior sentiment information.\\
    \item \underline{Text-Based Ideal Point (TBIP)} \cite{tbip}, an unsupervised Poisson factorisation model which can infer latent brand sentiment scores.\\
    \item 
 \underline{Brand Topic Model (BTM)} \cite{ZhaoGPH21}, a supervised Poisson factorisation model extended from TBIP with the incorporation of document-level sentiment labels.\\
    \item 
\underline{dBTM}, our proposed dynamic Brand Topic model in which the model is trained with the document-level sentiment labels at each time slice.\\

\item  \underline{O-dBTM}, a variant of our model that is only trained with the supervised review-level sentiment labels in the first time slice (denoted as the $0$-th time slice). In the subsequent time slices, it is trained under the unsupervised setting. In such a case, we no longer have a gold-standard brand ranking in time slices other than the $0$-th one. Instead of directly calculating the Spearman's rank correlation coefficient, we measure the difference of the brand ranking results in neighbouring time slices and use it to set the weight $\gamma^t$ in Eq. (\ref{eq:gamma}).
\end{itemize}

\paragraph{\textbf{Parameter setting}}
Frequent bigrams and trigrams\footnote{Frequent but less informative $n$-grams such as `\emph{actually bought}' were filtered out using NLTK.} are added as features in addition to unigrams for document representations. 
In our experiments, we train the models using the data from the current time slice and test the model performance on the full data from the next time slice. During training, we set aside 10\% of data in each time slice as the validation set. For hyperparameters, we set the batch size to 256, the maximum training steps to 50,000, the topic number to 50\footnote{The topic number is set empirically based on the validation set in the $0$-th time slice.}. 
It is worth noting that since topic dynamics are not explicitly modelled in the static models such as TBIP and BTM, their topics extracted in different time slices are not directly linked.

\section{Experimental Results}

In this section, we present the experimental results in comparison with the baseline models in brand rating, topic coherence/uniqueness measures, and qualitative evaluation of generated topics. 
For fair comparison, baselines are trained based on all previous time slices and predict on the current time slice.

\begin{figure*}[h]
\centering
\includegraphics[width=0.7\linewidth]{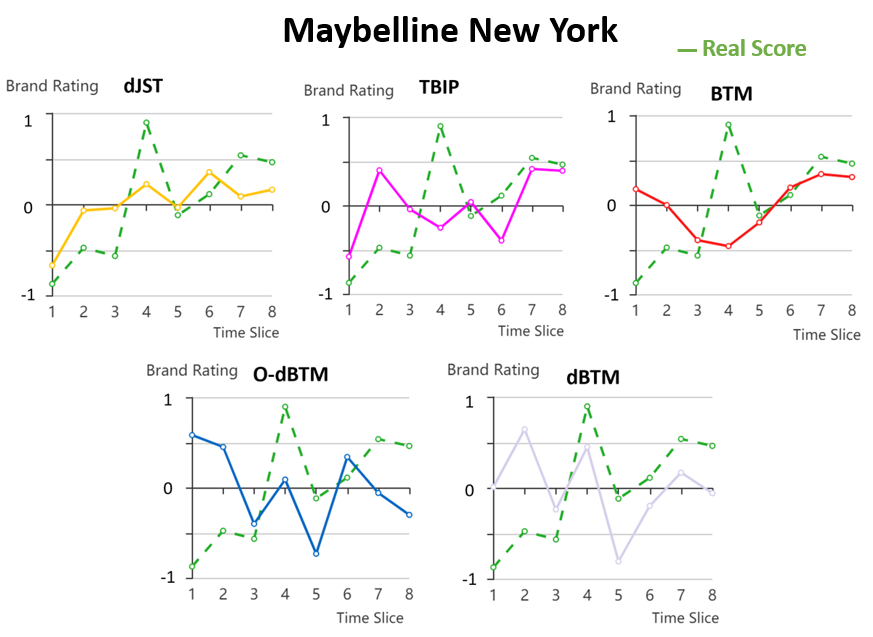}
\caption{The rating time series for `Maybeline New York'. The rating scores are normalised in the range of $[-1,1]$ with positive values denoting positive sentiment and negative ones for negative sentiment. In each subfigure, the dashed curve shows the actual rating scores.
}
\label{fig:ranking-series}
\end{figure*}

\subsection{Brand Rating}

TBIP, BTM and dBTM can infer each brand's associated polarity score automatically. For dJST, we derive the brand rating by aggregating the label distribution of its associated review documents and then normalising over the total number of brand-related reviews. The average of the document-level ratings of a brand $b$ at a time slice $t$ is used as the ground truth of the brand rating $x_b^t$. We evaluate two aspects of the brand ratings:

\paragraph{\textbf{Brand Ranking Results}}
We report in Table \ref{tab:brand-ranking} the brand ranking results measured by the Spearman's correlation coefficient, showing the correlation of predicted brand rating and the ground truth, along with the associated two-sided $p$-values of the Spearman's correlations 

Topic model variants, such as dJST, TBIP and BTM, produced brand ranking results either positively or negatively correlated with the true ranking results. We can see the correlation of BTM has switched between positive correlated and negative rated between time slices. With Gaussian state space models, our proposed model dBTM and its variant O-dBTM generate more stable results. \change{On MakeupAlley-Beauty}, dBTM gives the best results in 4 out of 8 time slices. Interestingly, O-dBTM with the supervised information supplied in only the first time slice outperforms the static models such as BTM in 3 out of 8 time slices, showing the effectiveness of our proposed architecture in tracking brand score dynamics. \change{Similar conclusions can be drawn on HotelRec that O-dBTM gives superior performance compared to BTM on 5 out of 6 time slices. Both O-dBTM and dBTM outperform the other baselines except TBIP in time slice 2.}

In summary, in dBTM, the brand rating score is treated as a latent variable (i.e., $x_{b_d}$ in Eq. \ref{eq:BTM-basic}) and is directly inferred from the data. 
On the contrary, models such as dJST, which require post-processing to derive brand rating scores by aggregating the document-level sentiment labels, are inferior to dBTM. This shows the advantage of our proposed dBTM over traditional dynamic topic models in brand ranking.

\paragraph{\textbf{Brand Rating Time Series}} 
The brand rating time series aims to compare the ability of models to track the trend of brand rating. For easy comparison, we normalise the ratings produced by each model, so that the plot only reflects the fluctuation of ratings over time. Figure~\ref{fig:ranking-series} shows the brand rating on the brand `Maybeline New York' generated on the test set \change{of MakeupAlley-Beauty} by various models across time slices. It can be observed that the brand ratings generated by TBIP and BTM do not correlate well with the actual rating scores. dJST shows a better aligned rating trend, but its prediction missed some short-term changes such as the peak of brand rating at time slice 7.
By contrast, dBTM correctly predicts the general trend of the brand rating. The weakly-supervised O-dBTM is able to follow the general trend but misses some short-term changes such as the upward trend from the time slice 1 to 2, and from the slice 6 to 7.

\begin{table*}[htb]
\centering
\resizebox{2\columnwidth}{!}{
\begin{tabular}{cccccccccccccccc}
\toprule
\multirow{2}{*}{Time Slice}  & \multicolumn{3}{c}{dJST}        & \multicolumn{3}{c}{TBIP} & \multicolumn{3}{c}{BTM}  & \multicolumn{3}{c}{O-dBTM}      & \multicolumn{3}{c}{dBTM}        \\
\cmidrule(lr){2-4} \cmidrule(lr){5-7} \cmidrule(lr){8-10} \cmidrule(lr){11-13} \cmidrule(lr){14-16} 
                  & coh    & uni   & quality        & coh    & uni   & quality & coh    & uni   & quality & coh    & uni   & quality        & coh    & uni   & quality        \\ \midrule
\multicolumn{16}{c}{\textit{MakeupAlley-Beauty}}      \\ \midrule
1          & -3.087    & 0.564      & 0.183  & -3.653    & 0.861      & 0.236& -3.836    & 0.862   & 0.225 & -3.486    & 0.820                        & \textbf{0.235} & -3.685    & 0.833      & 0.226          \\
2          & -3.008    & 0.513      & 0.170  & -4.043    & 0.850      & 0.210& -3.867    & 0.864   & 0.223 & -3.360    & 0.807                        & \textbf{0.240} & -3.642    & 0.829      & 0.228          \\
3          & -3.286    & 0.552      & 0.168  & -3.949    & 0.843      & 0.214& -3.716    & 0.851   & 0.229 & -3.369    & {\color[HTML]{333333} 0.787} & \textbf{0.234} & -3.611    & 0.823      & 0.228          \\
4          & -3.004    & 0.515      & 0.172  & -3.629    & 0.808      & 0.223& -3.837    & 0.846   & 0.220 & -3.457    & 0.771                        & 0.223          & -3.549    & 0.799      & \textbf{0.225} \\
5          & -3.112    & 0.560      & 0.180  & -4.168    & 0.838      & 0.201& -4.023    & 0.839   & 0.208 & -3.412    & 0.793                        & \textbf{0.232} & -3.523    & 0.818      & \textbf{0.232} \\
6          & -3.139    & 0.542      & 0.173  & -4.100    & 0.841      & 0.205& -3.976    & 0.846   & 0.213 & -3.433    & 0.761                        & 0.222          & -3.577    & 0.814      & \textbf{0.228} \\
7          & -3.269    & 0.521      & 0.159  & -4.049    & 0.854      & 0.211& -3.675    & 0.845   & 0.230 & -3.330    & 0.772                        & \textbf{0.232} & -3.667    & 0.825      & 0.225          \\
8          & -3.060    & 0.560      & 0.183  & -3.942    & 0.843      & 0.214& -3.715    & 0.837   & 0.225 & -3.589    & 0.789                        & 0.220          & -3.546    & 0.818      & \textbf{0.231} \\ \midrule
Average    & -3.120    & 0.541      & 0.173  & -3.942    & 0.842      & 0.214& -3.831    & 0.849      & 0.222 & -3.430    & 0.788                        & \textbf{0.230} & -3.600    & 0.820      & 0.228 \\ \midrule
\multicolumn{16}{c}{\change{\textit{HotelRec}}}      \\ \midrule
1       & -3.749 & 0.615 & 0.164 & -4.024 & 0.767 & 0.191 & -3.935 & 0.851 & 0.216 & -4.051 & 0.812 & 0.201          & -3.716 & 0.818 & \textbf{0.220} \\
2       & -4.020 & 0.633 & 0.158 & -3.577 & 0.753 & 0.211 & -3.960 & 0.813 & 0.205 & -3.851 & 0.803 & 0.209          & -3.696 & 0.809 & \textbf{0.219} \\
3       & -3.667 & 0.593 & 0.162 & -3.905 & 0.817 & 0.209 & -4.078 & 0.844 & 0.207 & -3.861 & 0.819 & 0.212          & -3.854 & 0.820 & \textbf{0.213} \\
4       & -4.008 & 0.644 & 0.161 & -3.747 & 0.808 & 0.216 & -3.946 & 0.859 & 0.218 & -3.637 & 0.814 & \textbf{0.224} & -3.681 & 0.794 & 0.216          \\
5       & -3.751 & 0.691 & 0.184 & -4.057 & 0.800 & 0.197 & -3.953 & 0.823 & 0.208 & -3.705 & 0.804 & 0.217          & -3.547 & 0.817 & \textbf{0.230} \\
6       & -3.916 & 0.697 & 0.178 & -3.770 & 0.810 & 0.215 & -4.061 & 0.855 & 0.210 & -3.510 & 0.800 & \textbf{0.228} & -3.705 & 0.821 & 0.222          \\ \midrule
Average & -3.852 & 0.645 & 0.168 & -3.847 & 0.793 & 0.206 & -3.989 & 0.841 & 0.211 & -3.769 & 0.809 & 0.215          & -3.700 & 0.813 & \textbf{0.220} 

\\\bottomrule
\end{tabular}
}
\caption{Topic coherence (\texttt{coh}) and uniqueness (\texttt{uni}) measures of the results generated by various models. We also combine the two scores to derive the overall \texttt{quality} of the extracted topics.}
\label{tab:brand-coherence-all}
\end{table*}

\begin{figure*}[t]
\centering
\includegraphics[width=\linewidth]{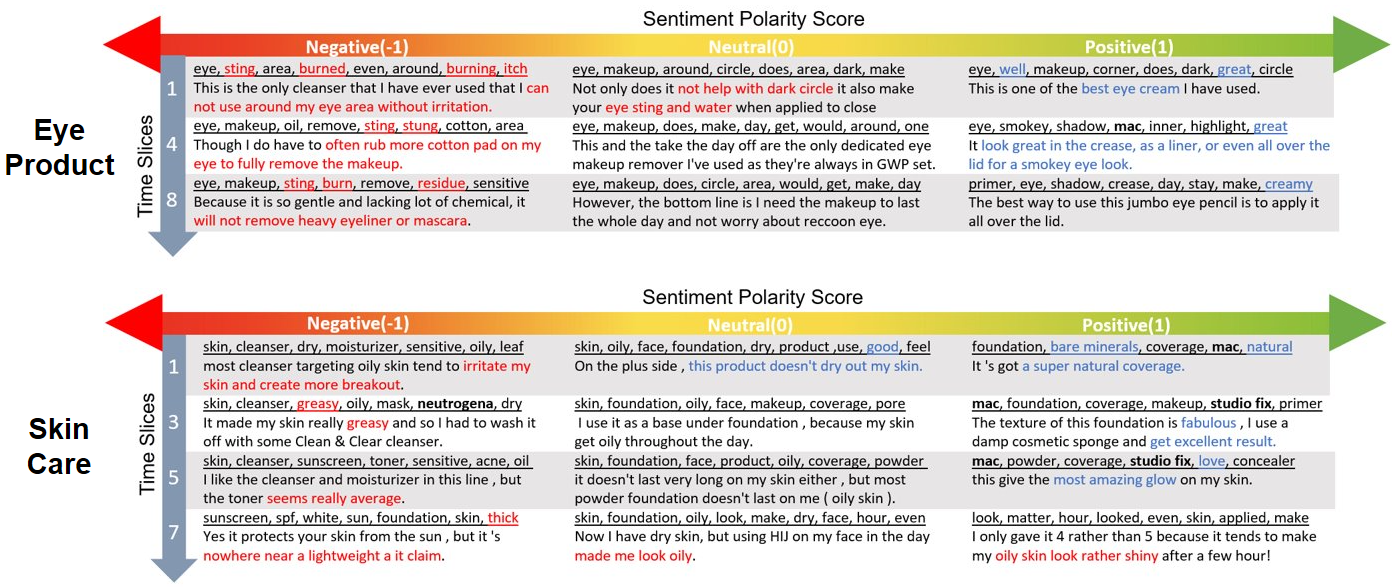}
\caption{Example of generated topics shown as a list of top associated words (underlined) in different time slices from the MakeupAlley dataset. 
For easy inspection, we also show the most representative sentence under each topic. The negative, neutral and positive topics in each time slice are generated by varying the brand polarity score from -1 to 0, and to 1. Positive words/phrases are highlighted in blue, negative words/phrases are in red, while brand names are in bold. 
}
\label{fig:topic-example}
\end{figure*}

\begin{figure*}[h!]
\centering
\includegraphics[width=\linewidth]{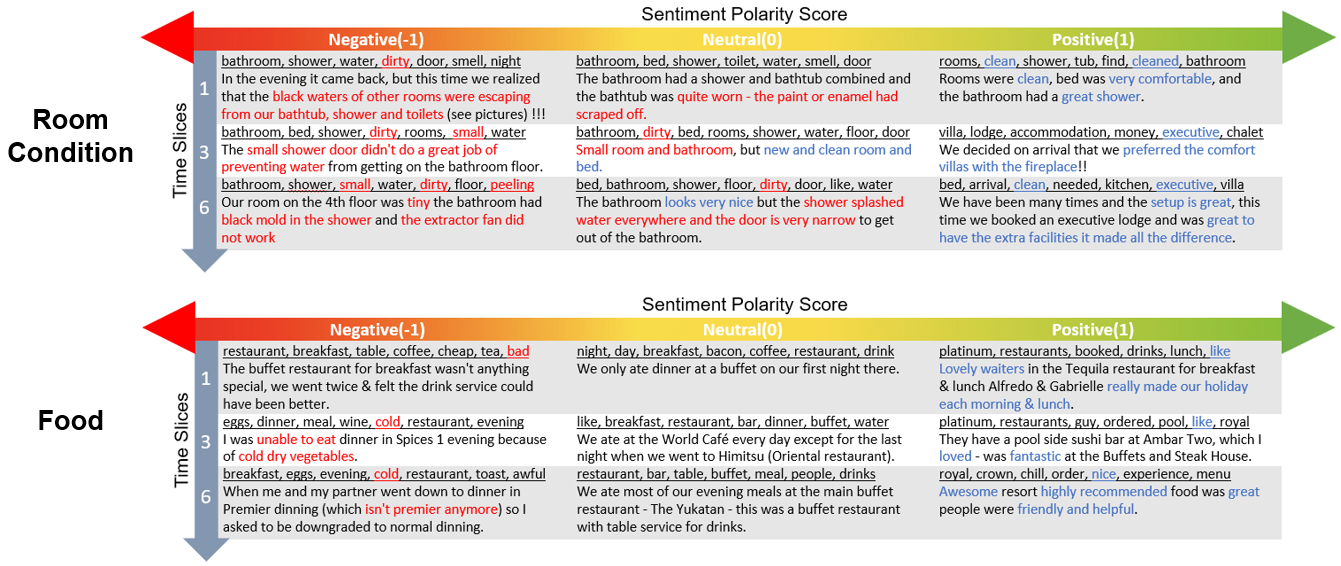}
\caption{\change{Example of generated topics shown as a list of top associated words (underlined) in different time slices from the HotelRec dataset. The representative sentence for each topic is also shown for easy inspection.}}
\label{fig:topic-example-hotel}
\end{figure*}

\subsection{Topic Evaluation Results}
We use the top 10 words of each topic to calculate the context-vector-based topic coherence scores \cite{RoderBH15} as well as topic uniqueness \cite{tmwa2019} which measures the ratio of word overlap across topics. We want to achieve balanced topic coherence and diversity. As such, topic coherence and topic diversity are combined to give an overall quality measure of topics \cite{tm2020}.
Since the results for topic coherence is negative in our experiment, i.e., smaller absolute values are better, we define the overall quality of a topic as $q=\frac{topic\ uniqueness}{|topic\ coherence|}$. Table~\ref{tab:brand-coherence-all} shows the topic evaluation results. 
In general, there is a trade-off between topic coherence and topic diversity. On average, dJST has the highest coherence but the lowest uniqueness scores, while TBIP has quite high uniqueness but the lowest coherence values. Both O-dBTM and dBTM achieve a good balance between coherence and uniqueness and outperform other models in overall quality.

\subsection{Example Topics across Time Periods}
We illustrate some representative topics generated by dBTM in various time slices. For easy inspection, we retrieve a representative sentence from the corpus for each topic.  For a sentence, we derive its representation by averaging the GloVe embeddings of its constituent words. For a topic, we also average the GloVe embeddings of its associated top words, but weighted by the topic-word probabilities. The sentence with the highest cosine similarity is selected.

Example of generated topics relating to `\emph{Eye Products}' and `\emph{Skin Care}' from MakeupAlley-Beauty is shown in Figure~\ref{fig:topic-example}.
We can observe that for the topic `\emph{Eye Products}', the top words of negative comments for `\emph{eye cleanser}' evolve from the reaction of skin (e.g. `\emph{sting}', `\emph{burned}') to the cleaning ability (e.g. `\emph{remove}', `\emph{residue}'). 
We could also see that the positive topics gradually change from praising the ability of the product for `\emph{dark circle}' in time slice 1 to the quality of eye shadow in time slice 4 and eye primer in time slice 8. Moreover, we observe the brand name \emph{M.A.C.} in the positive topic in time slice 4, which aligns with its ground truth rating. 
For the topic `\emph{Skin Care}', it can be observed that negative topics gradually move from the complaint of a skin cleanser to the thickness of a sunscreen, while positive topics are about the praise of the coverage of the M.A.C foundation more consistently over time. The results show that dBTM can generate well-separated polarity-bearing topics and it also allows the tracking of topic changes over time.

\change{Example of generated topics relating to `\emph{Room Condition}' and `\emph{Food}' from HotelRec is shown in Figure~\ref{fig:topic-example-hotel}. We can see that for the topic `\emph{Room Condition}', top words gradually shift from the expression of cleanliness (e.g. `\emph{clean}' in positive and `\emph{dirty}' in negative comments) to the description of the type and size of the rooms (e.g. `\emph{executive}' and `\emph{villa}' in positive reviews, and the concern of `\emph{small}' room size in negative comments). For the topic `\emph{Food}', the concerned food changes across time from drinks (e.g. `\emph{coffee}', `\emph{tea}') to meals (e.g. `\emph{eggs}', `\emph{toast}'). Negative reviews mainly focus on the concern of food quality, 
(e.g. `\emph{cold}'), 
while positive reviews contain a general praise of food and services (e.g. `\emph{like}', `\emph{nice}'). 
}

\subsection{Ablation Study}
We investigate the contribution of the meta learning component (i.e., Eq. 8 and 9) by conducting an ablation study and the results are shown in Table~\ref{tab:ablation}. We can observe that in general, removing meta learning leads to a significant reduction in brand ranking correlations across all time slices for the MakeupAlley-Beauty dataset. In terms of topic quality, we observe reduced coherence scores, but slightly increased uniqueness scores without meta learning, leading to an overall reduction of topic quality scores in most time slices. 

\begin{table}[h!]
\centering
\resizebox{\columnwidth}{!}{
\begin{tabular}{ccccccccc}
\toprule
\multirow{2}{*}{Time   Slice} & \multicolumn{4}{c}{dBTM}                                           & \multicolumn{4}{c}{dBTM (no meta learniing)}                    \\ \cmidrule(lr){2-5} \cmidrule(lr){6-9}
                              & cor            & coh             & uni            & quality          & cor               & coh       & uni                  & quality                \\ \midrule
\multicolumn{9}{c}{\textit{MakeupAlley-Beauty}}\\ \midrule
1                             & 0.402          & -3.685 & 0.833 & \textbf{0.226} & \textbf{0.435} & -3.972 & 0.861 & 0.217          \\
2                             & \textbf{0.438} & -3.642 & 0.829 & \textbf{0.228} & 0.189          & -3.704 & 0.840 & 0.227          \\
3                             & \textbf{0.523} & -3.611 & 0.823 & \textbf{0.228} & -0.162         & -3.873 & 0.828 & 0.214          \\
4                             & \textbf{0.453} & -3.549 & 0.799 & 0.225          & -0.042         & -3.745 & 0.849 & \textbf{0.227} \\
5                             & \textbf{0.394} & -3.523 & 0.818 & \textbf{0.232} & 0.086          & -3.990 & 0.832 & 0.209          \\
6                             & \textbf{0.433} & -3.577 & 0.814 & \textbf{0.228} & -0.029         & -3.958 & 0.856 & 0.216          \\
7                             & \textbf{0.402} & -3.667 & 0.825 & 0.225          & -0.042         & -3.587 & 0.842 & \textbf{0.235} \\
8                             & \textbf{0.364} & -3.546 & 0.818 & \textbf{0.231} & -0.125         & -3.920 & 0.847 & 0.216          \\ \midrule
\multicolumn{9}{c}{\change{\textit{HotelRec}}}                                                                                                                                            \\ \midrule
1                             & \textbf{0.285} & -3.716 & 0.818 & 0.220          & 0.222          & -3.559 & 0.791 & \textbf{0.222} \\
2                             & \textbf{0.382} & -3.696 & 0.809 & \textbf{0.219} & 0.210          & -3.796 & 0.801 & 0.211          \\
3                             & \textbf{0.355} & -3.854 & 0.820 & \textbf{0.213} & 0.285          & -3.763 & 0.790 & 0.210          \\
4                             & 0.315          & -3.681 & 0.794 & 0.216          & \textbf{0.408} & -3.597 & 0.817 & \textbf{0.227} \\
5                             & \textbf{0.364} & -3.547 & 0.817 & \textbf{0.230} & 0.362          & -3.657 & 0.793 & 0.217          \\
6                             & \textbf{0.312} & -3.705 & 0.821 & \textbf{0.222} & 0.262          & -3.694 & 0.809 & 0.219       \\
\bottomrule
\end{tabular}}
\caption{Results of dBTM with and without the meta learning component.}
\label{tab:ablation}
\end{table}

For HotelRec, we can see that removing meta learning also leads to a reduction in brand ranking results, but the impact is smaller compared to MakeupAlley-Beauty. For topic quality, we observe increased coherence but worse uniqueness, resulting in slightly worse topic quality results without meta learning in most time slices. One main reason is that unlike makeup brands where new products are introduced over time, leading to the change of discussed topics in reviews, the topic-word distribution does not change much across different time slices for hotel reviews. Therefore, the results are less impacted with or without meta learning.

\subsection{Training Time Complexity}

\begin{figure}[h!]
\centering
\includegraphics[width=0.7\linewidth]{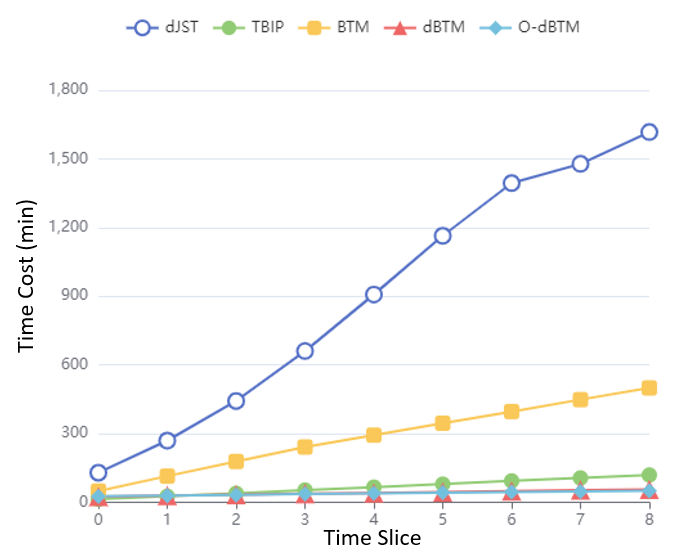}
\caption{Training time of models across time slices.}
\label{fig:running-time}
\end{figure}

All experiments were run on a single GeForce 1080 GPU with 11GB memory. 
The training time for each model across time slices is shown in Figure~\ref{fig:running-time}.  It can be observed that with the increasing number of time slices, the training time of dJST and BTM grows quickly. 
Both TBIP and dBTM take significantly less time to train. TBIP simply performs Poisson factorisation independently in each time slice and fails to track topic/sentiment changes over time. On the contrary, our proposed dBTM and O-dBTM are able to monitor topic/sentiment evolvement and yet take even less time to train compared to TBIP. One main reason is that dBTM and O-dBTM can automatically adjust the number of iterations with our proposed meta learning and hence can be trained more efficiently.

\section{Conclusion}
We have presented dBTM, which is able to automatically detect and track brand-associated topics and sentiment scores. 
Experimental evaluation based on the reviews from MakeupAlley \change{and HotelRec} demonstrates the superiority of dBTM over previous models in 
brand ranking and dynamic topic extraction. The variant of dBTM, O-dBTM, trained with document-level sentiment labels in the first time slice only, outperforms baselines in brand ranking and achieves the best overall result in topic quality evaluation. This shows the effectiveness of the proposed architecture in modelling the evolution of brand scores and topics across time intervals.

Our model currently only considers review ratings, 
but real-world applications potentially involve additional factors (e.g., 
user preference). A possible solution is to explore simultaneous modelling of user preferences to extract personalised brand polarity topics. 

\section*{Acknowledgements}

This work was supported in part by the UK Engineering and Physical Sciences Research Council (grant no. EP/T017112/1, EP/V048597/1, EP/X019063/1). YH is supported by a Turing AI Fellowship funded by the UK Research and Innovation (grant no. EP/V020579/1).

\bibliography{tacl2021}
\bibliographystyle{acl_natbib}

\iftaclpubformat

\onecolumn

\fi

\end{document}